\documentclass[10pt]{iopart}
\usepackage{epsfig}  
\begin{document}

\title[]{Energy Spectrum and Mass Composition around the Knee 
by EAS Measurements\footnote{talk given at the 18$^{\rm th}$ 
European Cosmic Ray Symposium, Moscow, July 2002.}}

\author{Andreas~Haungs\dag\ \footnote{To
whom correspondence should be addressed (Andreas.Haungs@ik.fzk.de)}
}

\address{\dag\ Forschungszentrum Karlsruhe, 
Institut f\"ur Kernphysik, D-76021 Karlsruhe, Germany}

\begin{abstract}
Primary cosmic ray particles above energies of about 100 TeV are
investigated by large-area ground based detector installations,
observing various components of the development of extensive 
air-showers (EAS). By such an indirect access to the primary 
cosmic ray spectrum a steepening of the power-law falls off at 
around 3-5 PeV, known as the {\em knee} has been identified.
Different experimental techniques are used to deduce mass- and 
energy-sensitive observables of EAS experiments.
The experiments involve measurements of secondary particle
distributions at various observation levels and of muons deep 
underground as well as of observables reconstructed by measuring 
the air Cherenkov light emitted with the shower development in the 
atmosphere. Recently methods for analysing multidimensional 
distributions get favoured since they are able to take
into account the correlations between different observables and the
influence of large intrinsic fluctuations of the air-shower 
development. Additionally the use of a larger set of observables
provides the possibility to test simultaneously the consistency of 
the Monte-Carlo models underlying the reconstruction procedures. 
By many experiments the existence of the knee in the primary energy 
spectrum is confirmed and a tendency of an increasing mean mass 
above the knee energy is indicated. Recent results show that the 
knee originates from a decrease of the flux of light primary 
particles, whereas heavy cosmic ray particles seem
to miss a kink in the energy range of 1 and 10 PeV.
\end{abstract}


\submitto{\jpg}


\section{Introduction}
The {\em knee} of the primary cosmic ray (CR) energy spectrum
has been discovered by Khristiansen and Kulikov~\cite{knee} in 
the year 1958. In their first conclusion the discoverers claimed 
that the kink in the spectrum is a consequence of the superposition 
of cosmic rays of galactic and supergalactic origin. Two years later 
B.~Peters pointed out with a theoretical argumentation that the 
position of the knee should get shifted in proportion to the charge 
of the primaries if magnetic fields would be responsible
for the acceleration of the CR~\cite{peters}.
For the following 40 years a lot of experiments~\cite{kalm} confirmed 
the kink in the energy spectrum (i.e. the change of the power-law 
index), displayed by various observables mapping the primary 
spectrum. Also astrophysical models accounting for such a kink got 
more detailed.
Nevertheless the key questions of the origin of the knee are still 
not convincingly solved. The ideas vary from astrophysical
reasons like rigidity cut-off at the 
acceleration~\cite{berezhko,biermann}, rigidity dependent escape 
from our galaxy~\cite{gaisser,puzhkin}, or the single-source model 
predicting a dominance of the flux at the knee by particles coming 
from a nearby supernova~\cite{erlwolf} to a scenario adopting a new 
channel of the primaries interaction producing new unobserved 
particles carrying away some energy~\cite{nikolsky,wigmans,kazanas}.
Many of these ideas and conjectures predict a detailed shape of the 
primary cosmic ray spectrum around the knee with specific variations 
of the elemental composition. Thus the experimental access to 
attack the basic questions are detailed and accurate measurements 
of the energy spectra of the various cosmic ray primaries, 
which imply also a determination of the energy variation of the 
CR mass composition.

It should be emphasised that all measurements around the knee are 
only of indirect character, i.e. the observation of primary CR is 
based on the interpretation and reconstruction of extensive 
air-showers (EAS). These techniques require a good knowledge of 
the shower development in the atmosphere and of the interaction 
mechanisms of high-energy particles with air-molecules. 
Extensive Monte Carlo simulation procedures are used for these 
purposes. For the high-energy hadronic interactions
more or less bold extrapolations from lower energies, formulated as
theoretical models and parameterizations, are at disposal. 
For the knee-energies accelerator data are not yet available 
(though the Tevatron collider is now close to the knee-energies) 
neither for relevant target-projectile combinations nor for the 
kinematic region of extreme forward interactions.
This situation leads to an uncertainty of unknown order, in some 
sense only to be estimated by using same reconstruction procedures 
based on the different hadronic interaction models.
Though in the last years the improved understanding of the 
high-energy hadronic interaction has reduced larger disagreements 
in predictions of air-shower observables~\cite{knapp},
it remains unclear if there are common systematic uncertainties 
in the interaction models by unknown features and interaction paths, 
not yet considered.
Hence results of air-shower experiments around the knee region may 
be considered in three categories: 
i) Information on the primary energy spectrum and chemical 
composition in absolute scale. 
ii) Phenomenological information with a minimised influence of the 
Monte Carlo simulations, but with general results for the 
understanding of the origin of the knee. 
iii) Proofs and tests of Monte Carlo models by using redundant 
information of the experimental investigations. \\
The concern to the latter aspect has been emphasised in last years 
by invoking the possibilities of multi-detector 
setups~\cite{antoni99,antoni01}.
The validity of the hadronic interaction models used as generators 
of Monte Carlo simulations has got an important subject in context 
of EAS analyses. Here a co-operation between present and future 
accelerator experiments and the cosmic ray investigations is aspired. 
Some recent initiatives have been started to establish such 
interrelated efforts, e.g. the NEEDS (Needs from Accelerator 
Experiments for the Understanding of High-Energy Extensive 
Air-Showers) workshop, held in Forschungszentrum Karlsruhe in April 
2002~\cite{needs}, where physicists of both the high-energy 
accelerator and the cosmic ray communities commonly discussed about 
future possibilities of mutual assistance and exchange of 
relevant data.

\section{The all-particle cosmic ray energy spectrum}
\begin{figure}[ht]
\begin{center}
\vspace*{-.1cm}
   \epsfig{file=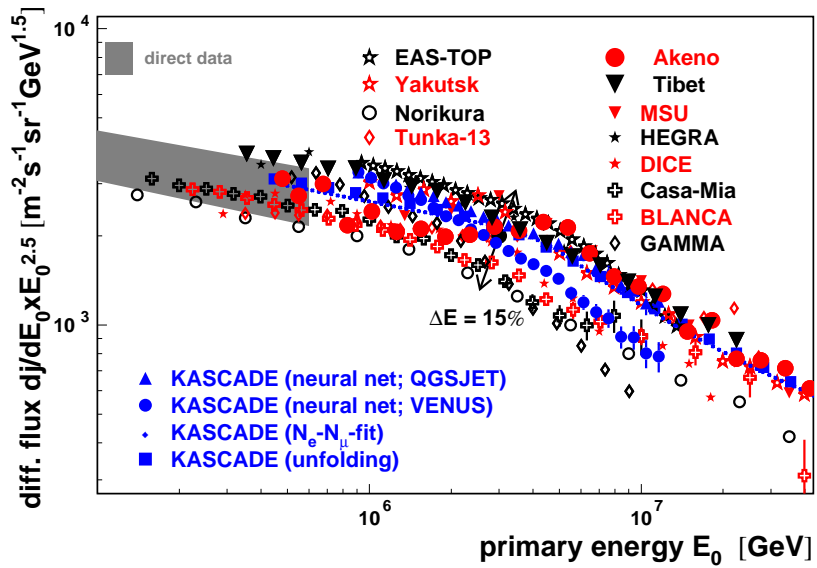,width=11.cm}
\vspace*{-.6cm}
\end{center}
\caption{\label{worldE}
Compilation of different experimental results on the cosmic ray 
all-particle energy spectrum around the knee. The effect of a 
15\% uncertainty in energy reconstruction is indicated.
(Refs.: Akeno~\cite{akeno}, Tibet~\cite{tibet}, MSU~\cite{msu}, 
HEGRA~\cite{hegra}, DICE~\cite{dice}, Casa-Mia~\cite{casa-mia},
Blanca~\cite{blanca}, Gamma~\cite{gamma}, EAS-TOP~\cite{eastop}, 
Yakutsk~\cite{yakutsk}, Norikura~\cite{norikura}, 
Tunka~\cite{tunka}, KASCADE~\cite{kas-roth,ralph,ulrich},
compilation of direct data by Watson~\cite{watson}.)}
\vspace*{-.1cm}
\end{figure}

The reconstruction of the all-particle energy spectrum from 
air-shower measurements was proceeded over many years with basically 
similar experimental devices and procedures: An array of detectors 
spread over an area of several thousands of square-meters measures 
particle and/or Cherenkov light densities. The lateral density 
distributions and integral quantities of them are 
analysed (adopting some experiences gained by Monte Carlo 
simulations) to infer the primary energy of the incident particles, 
either for single events or more often for the average of the 
measured distributions. These procedures require a-priori 
assumptions of the elemental composition, since the lateral 
distributions and total intensities differ for different primary 
masses of the shower inducing particle.
The idea is to realise a calorimetric determination of the energy 
of shower cascades by the total Cherenkov light intensity produced in 
the atmosphere or by the total number of (charged) particles at a 
certain observation level.
The determined quantities are assumed to be proportional to the 
primary energy. \\
Typical examples of such measurements are presented by  
the energy spectra obtained by the Akeno~\cite{akeno}
and Tibet~\cite{tibet} detector arrays. Both experiments use a 
field of scintillation detectors to measure the charged particle 
component. Akeno converted the obtained size of the showers into 
the  primary energy by help of EAS results measured at 
observation levels of high altitude (conversion in N$_e^{\rm max}$)
and a theoretically anticipated parameter for the dependence of the 
shower size maximum from the primary energy. 
Tibet uses Monte Carlo simulations for fitting the obtained size 
spectrum introducing the elemental composition as free parameters.
The two energy spectra from the two different experiments 
(see Fig.~\ref{worldE}) differ by the position
and the sharpness of the knee: Akeno attained a much
sharper knee at an energy position higher than the results of 
the Tibet experiment show. Such differences are subject of current 
discussions. \\
Fig.\ref{worldE} compiles the energy spectra obtained by various 
different experiments. Most of them use for the reconstruction of 
the energy a single observable like the EAS size and convert the 
obtained size spectrum with help of Monte Carlo simulations.
In contrast, the multi-detector setup of the KASCADE experiment e.g.
enables to determine both the electron and muon sizes of each  
single EAS event. Various methods are  applied to reconstruct the 
primary energy spectra in a consistant way using both informations 
simultaneously. In this manner the correlated uncertainty on energy 
and mass gets disentangled.
A common fit to the electron and muon size spectra with the 
assumption of two primary mass group spectra following two power 
laws with a knee (8 free parameters: spectral indices 
below and above the knee,
the knee position, and the normalization factors for the light and
heavy spectra, respectively) leads to the energy spectra
of the two components and the all-particle energy spectrum as 
sum~\cite{ralph}. The fit is based on a Fredholm integral with 
the detector response function as kernel.
This response function is prepared by detailed Monte Carlo 
calculations including a full simulation of the detector properties.
A step forward is the application of an unfolding procedure on
the two size spectra~\cite{ulrich} leading to the mass group
spectra of four components. A further analysis using the two sizes 
per single event is based on neural net estimations of the primary
energy~\cite{kas-roth}. The latter procedure is also sensitive to 
the influence of different hadronic interaction models as the net 
is trained by two different models. \\
Despite of considerable differences of the applied methods 
analysing the observables, different simulation procedures, 
and different observation levels the compiled experimental results 
agree in a remarkable way.
Assuming an uncertainty of 15\% for the energy reconstruction
(shown in Fig.~\ref{worldE} at $3\cdot 10^{15}\,$eV) all the results
appear to be concordant with each other. Only at the high-energy 
ends the spectra exhibit somehow larger differences, may be as 
consequence of saturation effects of the different detector devices 
in addition to missing statistical accurracy and also due to 
larger uncertainties of the models providing the reference patterns.
This observation is remarkable, first of all as most of the 
experimental results have been published without quoting systematic 
uncertainties due to the reconstruction procedures or the model 
dependence. After compiling the data (14 spectra) the average values 
and their variances result in a slope below the knee to 
$\gamma_1 = -(2.68\pm0.06)$
and above the knee to $\gamma_2 = -(3.06\pm0.08)$ with the knee 
position at $E_k = (3.2\pm1.2)\cdot 10^{15}\,$eV, without taking 
into account the statistical accuracy of the different experiments 
or any further systematic uncertainty.

\section{The cosmic ray elemental composition}
Whereas the energy reconstruction can be based on one single 
measurable quantity, the determination of the mass composition 
needs at least a second observable, which characterises in relation 
to the first the actual status of the EAS development 
which is specific for the mass of the primary.
The first interaction of a cosmic nucleus of mass A with
an air molecule determines the further development of the air shower, 
and the mechanisms of this interaction are responsible for the 
formation of mass sensitive observables at earth-bound experiments.
\begin{figure}[ht]
\begin{center}
\vspace*{-.1cm}
   \epsfig{file=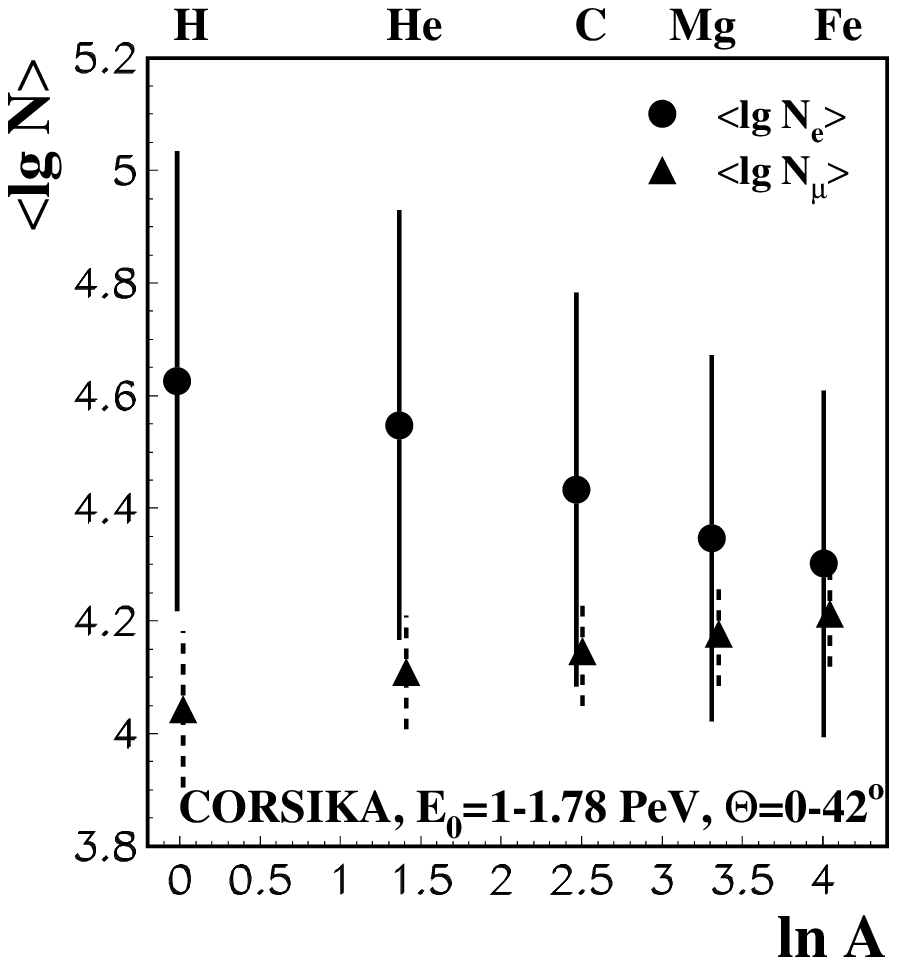,width=5.6cm}
   \epsfig{file=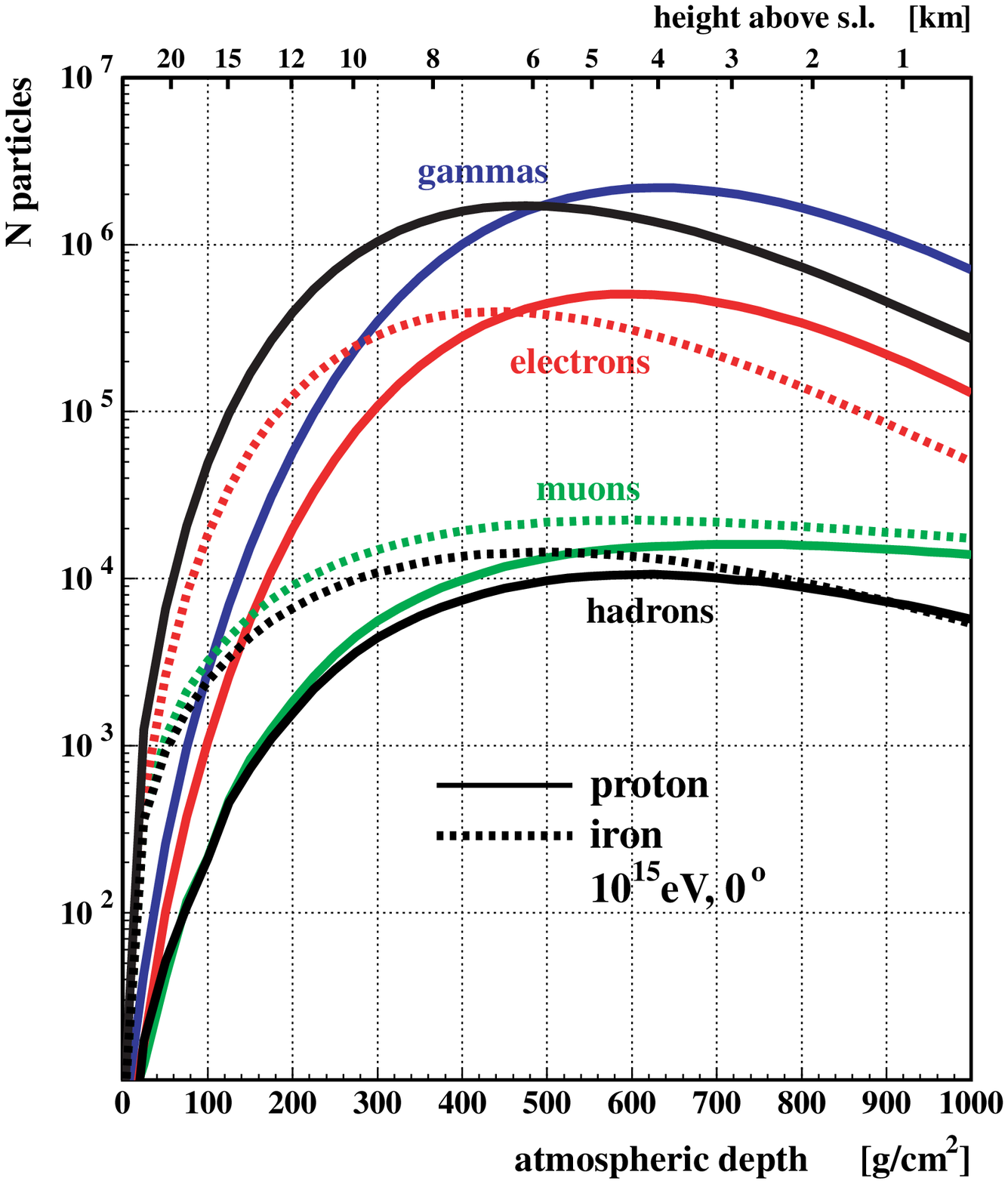,width=6.6cm}
\vspace*{-.5cm}
\end{center}
\caption{\label{princip}
Left panel: Dependence of the electron and muon particle 
numbers of EAS on the primary mass A. The error bars represent 
the fluctuations in the numbers (at sea-level). 
Right panel: Longitudinal development of the various particle
components of proton and iron induced EAS.
The simulations are done with CORSIKA/QGSJET~\cite{cors,qgs}.}
\vspace*{-.1cm}
\end{figure}
With increasing mass of the target or
projectile the cross sections of nucleus-nucleus
interactions get larger. For example, the inelastic cross section
$\sigma_{inel}^{A-Air}$ of iron is at $1\,$PeV approximately
6 times larger than for protons of equal energy.
Hence an EAS starts earlier in average and develops faster
in the atmosphere with increasing primary mass.
In a first good approximation a primary nucleus of mass A and
energy $E_0$ can be regarded for the shower development
as a swarm of A independent nucleons generating A superimposed
independent proton showers of the energy $E_0/A$
(superposition principle). 
As a consequence showers induced by heavy primaries
generate more secondary particles, each of  smaller  
energy, and due to the faster attenuation of 
the electromagnetic component 
\begin{figure}[ht]
\begin{center}
\vspace*{-.05cm}
   \epsfig{file=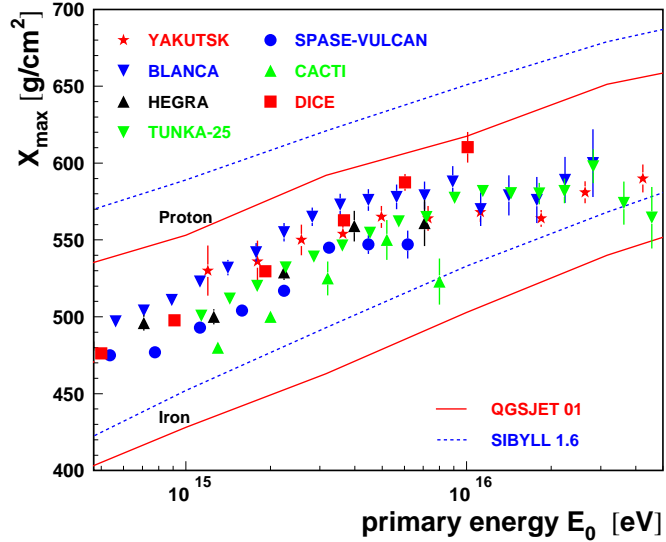,width=9.cm}
\vspace*{-.5cm}
\end{center}
\caption{\label{xmax}
Compilation of different experimental results on the estimation
of the shower maximum by measuring the Cherenkov light.
(Refs.:Yakutsk~\cite{yakutsk}, Blanca~\cite{blanca}, HEGRA~\cite{hegra},
Tunka-25~\cite{tunka25}, Spase~\cite{spase}, Cacti~\cite{cacti}, 
Dice~\cite{dice2}.)}
\vspace*{.1cm}
\end{figure}
with a smaller number of electrons at the observation level 
(after the EAS maximum).
Simultaneously the number of muons is larger. 
The muons interact weakly with the atmosphere, 
they are less absorbed and their decay time is long compared to 
pions, so that they add up throughout the shower development.
The superposition model predicts for all additive observables
power law dependences with the mass. Even if the principle is
an approximation, this dependence is sufficiently valid
(Fig.~\ref{princip}).
From statistical reasons the fluctuations of the sum of A 
independent showers should be smaller than of a shower generated 
by a single proton of higher energy (Fig.~\ref{princip}).
The effect is smeared out by the limits of the superposition
model in the interaction, but it remains efficient for
mass separation in experiments.
Further, surviving hadrons have less energy, and compared
to the electron or muon number the total number
of hadrons in EAS is small and they are concentrated around the
shower axes.
The faster development of showers induced by heavy primaries lead
\begin{figure}[ht]
\begin{center}
\vspace*{.2cm}
   \epsfig{file=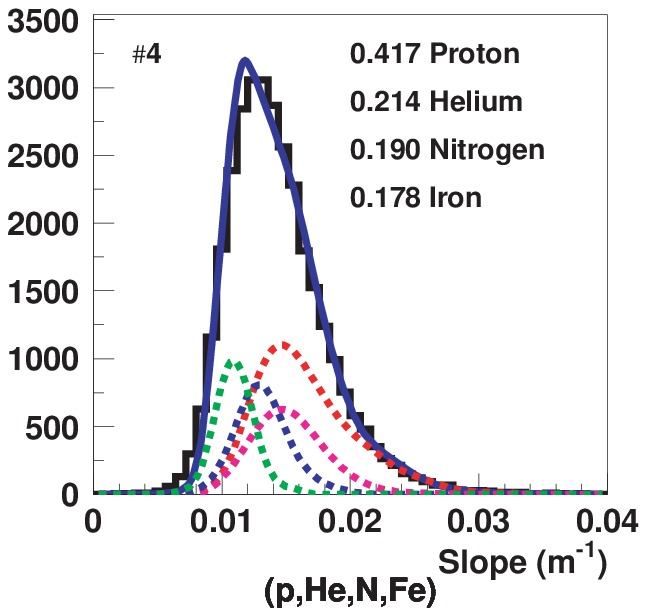,width=4.8cm}\hspace*{0.2cm}
   \epsfig{file=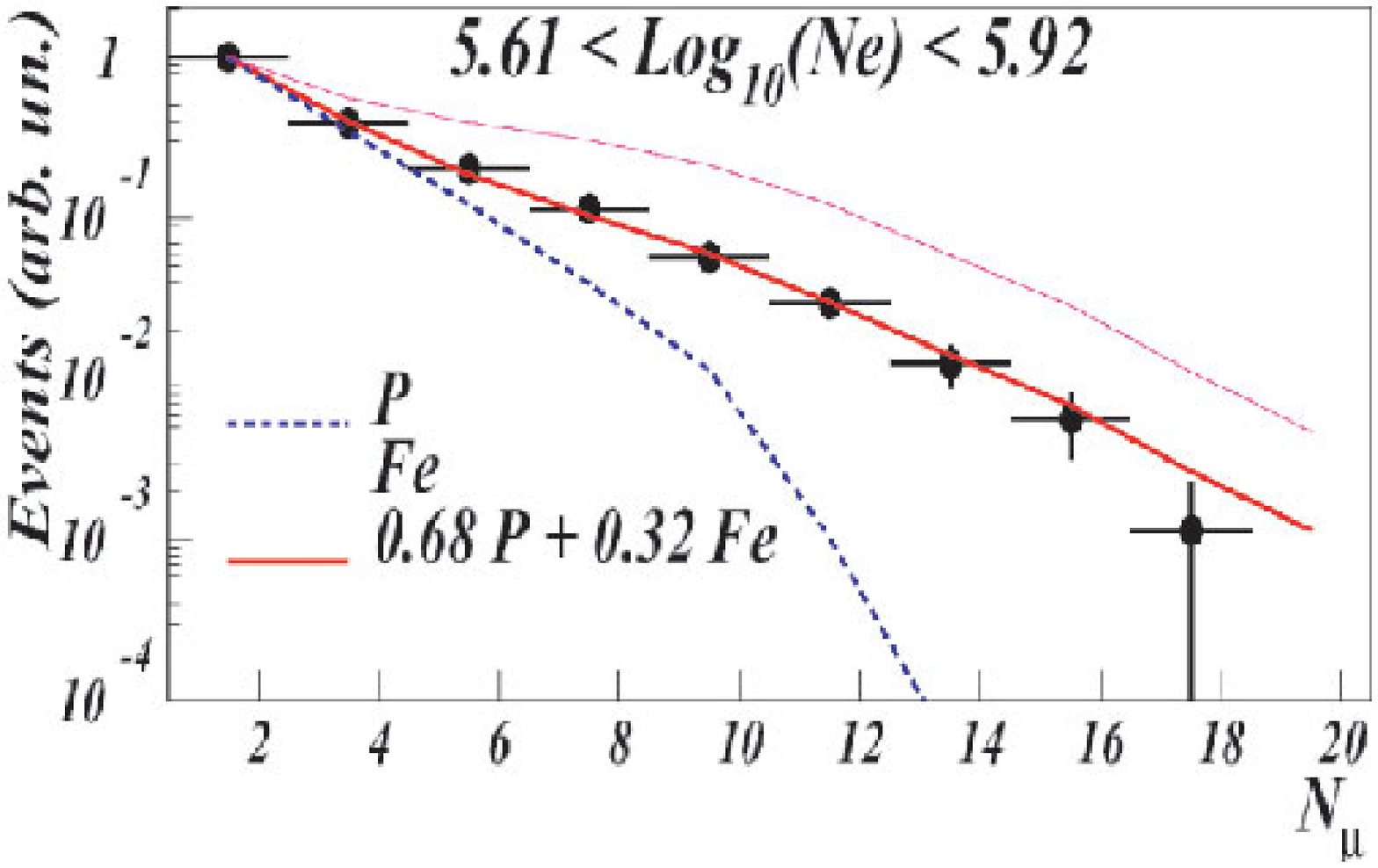,width=7.3cm}
\vspace*{-.5cm}
\end{center}
\caption{\label{comp-ex}
Distributions of mass sensitive observables 
(left panel: slope of the lateral distribution of the 
atmospheric Cherenkov light as measured with Blanca~\cite{blanca}
for the energy range of $10^{14.9}-10^{15.3}\,$eV,
right panel: muon multiplicity distribution measured with
MACRO~\cite{eas-macro} deep underground) compared with predictions
of Monte Carlo simulations for an energy range just below the knee.}
\vspace*{-.1cm}
\end{figure}
additionally to larger relative angles to the shower axes of 
the secondaries and flatter lateral distributions. \\
The atmospheric Cherenkov light, produced dominantly by relativistic
\begin{figure}[ht]
\begin{center}
\vspace*{0.1cm}
   \epsfig{file=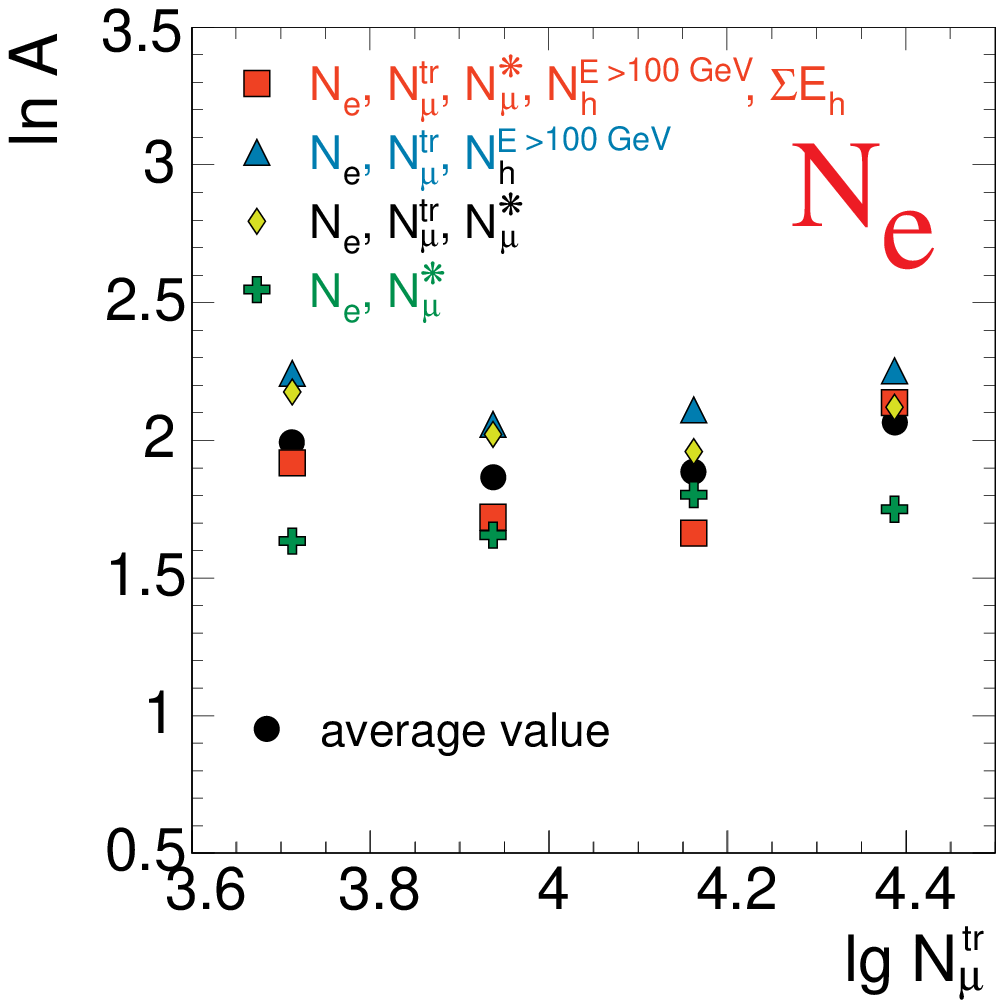,width=5.03cm}
   \epsfig{file=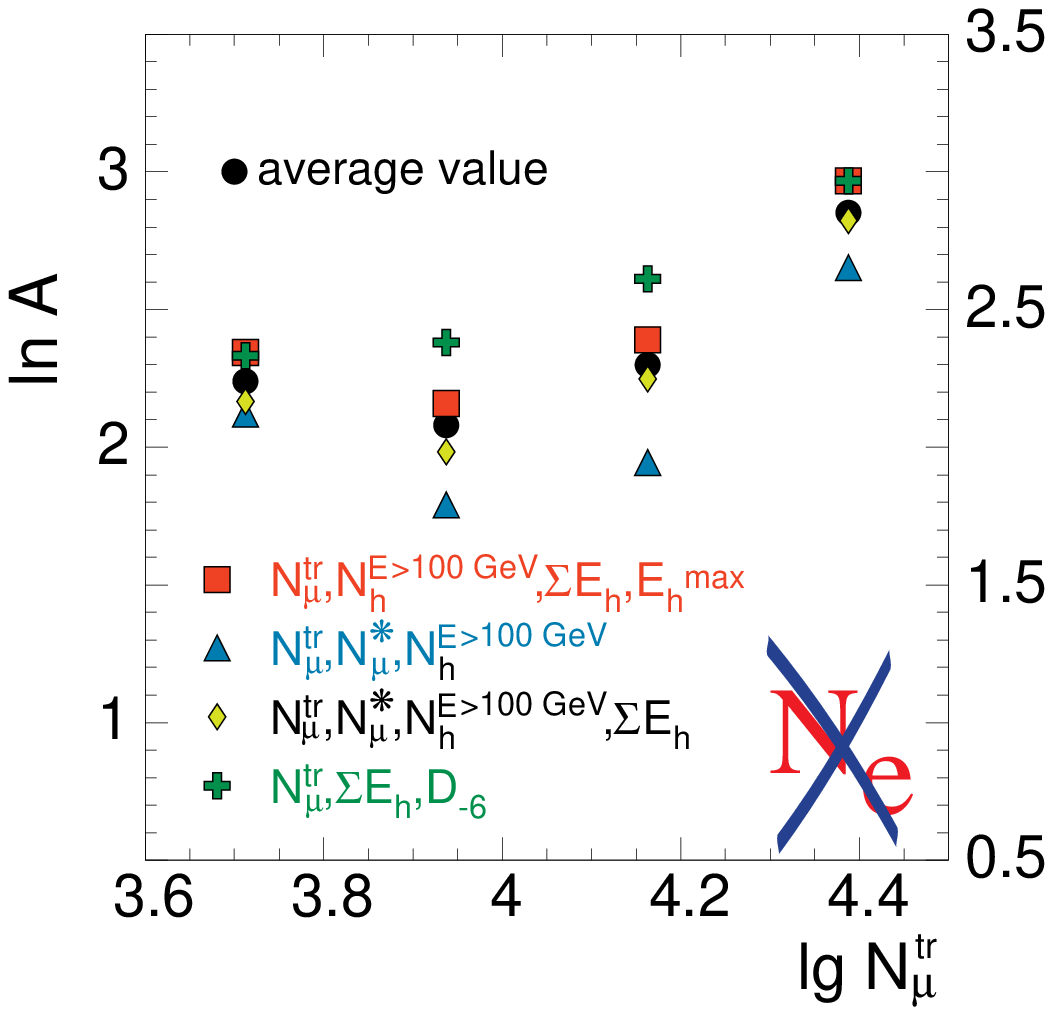,width=5.5cm}
\vspace*{-.5cm}
\end{center}
\caption{\label{comp}
Energy dependence ($\propto N_\mu^{tr}$) of the mean
logarithmic mass using different sets of observables
(shower sizes $N_e$, $N_\mu^{tr}$ as well as hadronic 
and high-energy muon observables
reconstructed at the central detector of KASCADE)
in a multivariate approach. Systematic differences appear when the
EAS electron size is included in the correlation or not 
(from~\cite{kas-roth}).}
\vspace*{.1cm}
\end{figure}
electrons and positrons is directly sensitive to the
height of the shower maximum, which is more distant from the
observation level for heavier primaries
(Figs.~\ref{princip} and~\ref{xmax}).
Further mass sensitive differences in the longitudinal shower 
development, but less pronounced and with difficulties to measure, 
are the relative arrival times of the secondary particles of the 
shower disc, in particular of the muons  or of the Cherenkov light 
and the relative angles of incidence of EAS muons. \\
In addition to different observables considered in various 
\begin{figure}[ht]
\begin{center}
\vspace*{-.1cm}
   \epsfig{file=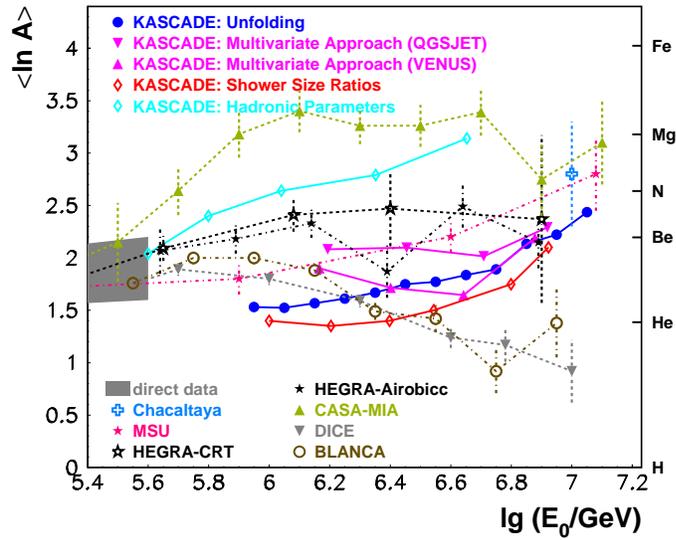,width=9.cm}
\vspace*{-.5cm}
\end{center}
\caption{\label{worldA}
Compilation of different experimental results of  the mean 
logarithmic mass around the knee. The lines are for guiding the eyes
and connect the data points of the experiments
(Chacaltaya~\cite{chaca}, MSU~\cite{msu}, HEGRA-CRT~\cite{crt},
HEGRA~\cite{hegra}, Casa-Mia~\cite{casa-comp}, Blanca~\cite{blanca},
Dice~\cite{dice}, KASCADE~\cite{ulrich,kas-roth,weber,hoer})}
\vspace*{-.1cm}
\end{figure}
experiments, also different reconstruction techniques are applied.
One approach implies the comparison of the distributions of the 
mass sensitive observables with Monte Carlo predictions for 
different primary masses in bins of energy.
Examples for that are the MSU~\cite{msu} experiment 
(fluctuations of muon numbers), Blanca~\cite{blanca} (slope of 
the lateral distribution of the Cherenkov light, Fig.~\ref{comp-ex}), 
EAS-TOP/MACRO~\cite{eas-macro} (muon multiplicity
measured deep underground, Fig.~\ref{comp-ex}), 
Chacaltaya~\cite{chaca}
(hadron multiplicity measured in burst chambers), or the ratio 
of muon to electron size like used in KASCADE~\cite{weber} or
CASA-MIA~\cite{casa-comp}. \\
Fig.\ref{xmax} compiles different results on the quantity
X$_{\rm max}$ of experiments sensitive to the Cherenkov light
and points out how sensitive the measurements are to the 
uncertainties. Up to 40 g/cm$^2$ difference for different 
high-energy interaction models, and up to 60 g/cm$^2$ variance 
due to sytematic experimental uncertainties in addition to the
uncertainty for the energy estimation may explain the scatter of 
the data points in the figure so that actually no definite 
conclusion is possible on the mass composition on basis of 
this observable. \\
Uncertainties in the model calculations for X$_{\rm max}$ 
are naturally transmitted to expectations of particle 
distributions at observation level. Consequently results
obtained by measuring charged particles also scatter 
enormously (Fig.~\ref{worldA}).
It was found by the KASCADE collaboration that results of the mean 
logarithmic mass depend strongly on the choice of 
observables~\cite{kas-roth}.
KASCADE is able to register a large set of observables for EAS 
hitting the Central Detector of KASCADE.
The same analyses method (neural net classification trained with 
always the same set of simulations) is applied to sets of 
different observables (Fig.~\ref{comp}).
Most of the results indicate an increase of the mean 
logarithmic mass above the knee.
The results point to an invalid balance of energy and particle 
numbers predicted by the interaction models. Averaged values of 
such a multivariate approach
are displayed for two different interaction models underlying the 
neural net training among some other experimental results at 
Fig.~\ref{worldA}.

\section{Behaviour of single mass groups}
A deeper insight in the structure of the knee is provided by the 
spectra of single mass groups. 
The KASCADE experiment e.g., measures with its multi-detector setup
local muon density spectra for different muon energy thresholds for
event samples of fixed core distances~\cite{kas-haungs}.
The event selection is performed
with help of the scintillator array, whereas the muon densities are
estimated independently by devices placed in the center of the array.
Moreover, the possibility to reconstruct the electron to muon number
\begin{figure}[ht]
\begin{center}
\vspace*{-.1cm}
   \epsfig{file=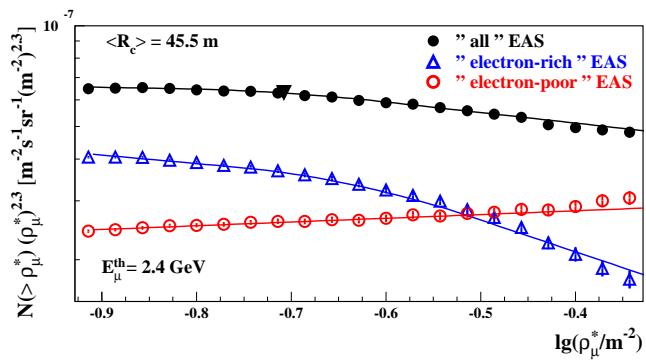,width=9.5cm}
\vspace*{-.5cm}
\end{center}
\caption{\label{mudens}
Spectra of the local muon density $\rho_\mu^\star$ for a fixed 
distance to the shower center $R_c$ for different subsamples of 
EAS as measured by the KASCADE experiment (from~\cite{kas-haungs}).
Only light induced showers (electron-rich) show a knee in the energy
range of 1 to 10 PeV.}
\vspace*{-.1cm}
\end{figure}
in each station of the array allows to divide the total sample of EAS
in an electron-rich and an electron-poor subsample. This selection
is nearly independent of Monte-Carlo assumptions, as it is well known
that showers induced by light primaries have a larger ratio than 
EAS induced by heavy primaries. The analysis results in the statement 
that the 
local muon density spectra (which reflect the primary energy spectra) 
show a knee in the total sample, a even more pronounced knee for 
the electron-rich (light primaries) sample, and no knee for the 
electron-poor (heavy primaries) sample (Fig.~\ref{mudens}). 
The accessible energy range for this analysis lies
between 1 and 10 PeV. This result was found to be valid for many 
\begin{figure}[ht]
\begin{center}
\vspace*{.1cm}
   \epsfig{file=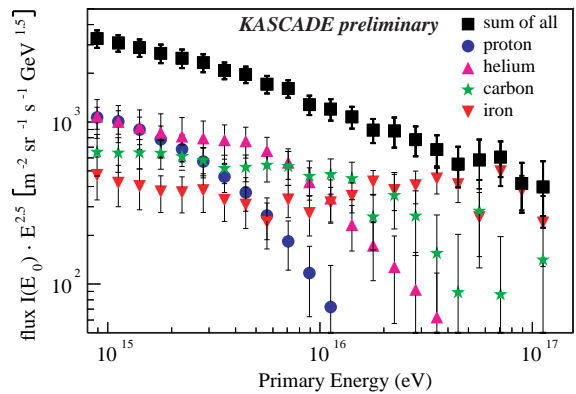,width=7.5cm}
\vspace*{-.5cm}
\end{center}
\caption{\label{unfold}
Energy spectra of four primary mass groups as obtained from an
unfolding procedure applied to the KASCADE size spectra
(from~\cite{kas-kampert}) The error bars display only the 
statistical uncertainties.}
\vspace*{-.1cm}
\end{figure}
different core distances and for two muon energy thresholds
and is a strong experimental hint for the origin of the knee from a
decrease of the flux of light primary particles. \\
Combining surface measurements of the shower size with underground 
muon multiplicities by the EAS-TOP and MACRO 
experiments also a decrease of the contribution of 
light primaries to the flux above the knee is 
observed~\cite{eas-macro}. \\
A more promising way towards a solution of the puzzle of the knee 
is a new approach worked out by the KASCADE collaboration: 
Large statistical accuracy and reliable estimates of the electron 
and muon size per single shower allow to apply unfolding methods 
to the two-dimensional size distribution obtaining the energy 
spectra of different mass groups.
Mathematically the inverse problem is put:
$g(y) = \int{R(y,x)p(x)d(x)}$ with $y=(N_e,N_\mu^{\rm tr})$
and $x=(E,A)$. The quantity $R(y,x)$ is the response matrix 
which includes additionally to the mass sensitive shower sizes 
also the intrinsic shower fluctuations, which are mass dependent, 
too. The response function has to be deduced from Monte Carlo 
simulations.
Different methods are applied to solve the problem: An iterative
procedure using the Gold-algorithm based on a minimization of the
$\chi^2$-function~\cite{ulrich-isv}. Alternatively
an approach based on the Bayes-theorem is applied~\cite{roth-isv},
where for each mass-energy-bin the probabilities have to be
calculated how the $N_e-N_\mu^{tr}$-bins will be populated.
An iterative procedure tries subsequently to
reproduce the given two-dimensional size-distribution of the data.
Whereas the kernel function for the first approach is calculated by
combining simulations of high stastical accuracy, using the 
thinning procedure and full simulations to describe all physical 
and experimental effects (fluctuations of shower size, efficiencies, 
reconstruction accuracies), the latter approach requires a larger 
statistical accuracy of full simulations, but does not require 
the inversion of the response matrix. \\
In a first step the unfolding method using the Gold-algorithm was 
applied to the one-dimensional spectra of the sizes, only. 
Preliminary results were reported by the KASCADE 
collaboration~\cite{ulrich,kas-kampert} and confirm again (but still
with large systematic uncertainties) that the 
knee is due to the spectrum of light primaries.
As a new aspect the positions of the knee for the different mass
groups suggest a rigidity dependent behaviour of the knee
(Fig.~\ref{unfold}). Though the absolute fluxes and slopes of the
different mass groups still have large uncertainties arising 
from the limited number of Monte Carlo simulations, from model 
uncertainties, and methodological reasons, the results indicate to 
the rigidity dependent knee position.

\section{Conclusions and future}
\begin{figure}[ht]
\begin{center}
\vspace*{-.01cm}
   \epsfig{file=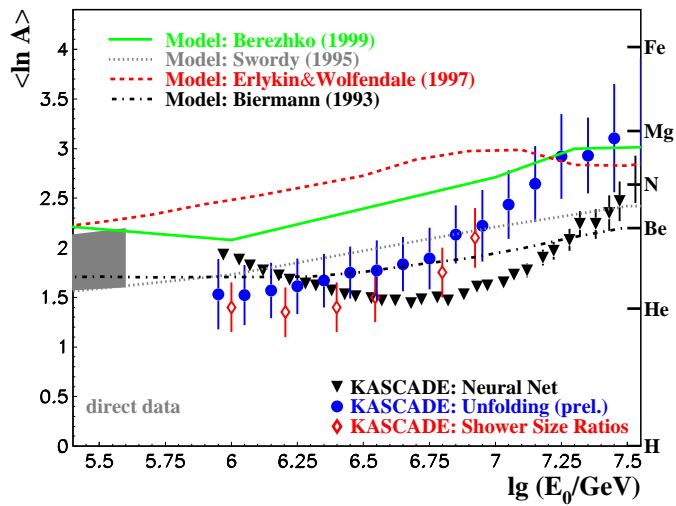,width=9.cm}
\vspace*{-.5cm}
\end{center}
\caption{\label{models}
The mean logarithmic mass distributions resulting from different 
analyses of the KASCADE experiment~\cite{kas-roth, ulrich, weber}) 
as compared with predictions of astrophysical models
for the origin of the knee (Refs.: Berezhko~\cite{berezhko}, 
Swordy~\cite{swordy},
Erlykin and Wolfendale~\cite{erlwolf}, Biermann~\cite{biermann}.)}
\vspace*{-.1cm}
\end{figure}

It remains still unclear if the present results are able to 
distinguish between different theoretical rigidity-dependent 
models for the origin of the knee.
Fig.~\ref{models} shows results of the mean
logarithmic mass deduced by 3 different methods applied for the
KASCADE experiment using identical data samples and the same 
high-energy interaction model generating the reference patterns
(QGSJET). 
Hence most sources of systematic uncertainties should be 
eliminated. Remaining differences in estimating mean mass and
energy must be of methodical origin. It is evident that the 
resulting uncertainties are of the same 
order as the differences of the astrophysical models.
The experimental results favour models predicting a lower mean 
logarithmic mass of the CR mass composition,
and they have obvious difficulties to match e.g. the model of
Erlykin and Wolfendale~\cite{erlwolf}
where the knee display a composition dominated by oxygen nuclei, 
with a steeply falling
proton component below the knee and leading to a large mean 
mass number.
Models with a simple
rigidity dependent knee and identical spectral indices 
before and after 
the knee (e.g. the Biermann-model~\cite{biermann}) 
tentatively better agree with the data. More definite conclusions, 
however, are presently hardly possible. \\
For a consolidation of the rigidity dependent knee position 
(i.e. ascribing the knee to an astrophysical origin and excluding 
features due to changements of the interaction scaling with the 
atomic mass) a kink for the heavy (iron) component at 
$\approx 10^{17}\,$eV should be observed. 
This is necessary in order to confirm the above stated results
and to match the total energy spectrum at higher energies. 
A large number of the present detector installations 
measure around the (proton) knee region or they are optimized for 
much higher energies around the Greisen-Zatsepsin-Kuzmin 
cutoff (GZK). None of the modern multidetector experiments is  
optimised for the energy range of $10^{16}$ to $10^{18}\,$eV.
Only the Yakutsk array~\cite{yakutsk} which registers air-Cherenkov
radiation from EAS covers the full energy range from knee region 
to GZK cut off energies.
The interest of newer EAS observation experiments 
(HESS, Argo, Milagro, e.g.) is directed to 
the detection of high-energy Gamma ray sources in the TeV region 
(in this way attacking also the origin of the charged primary 
cosmic rays - see G.Heinzelmann, these proceedings).
These experiments measure as background charged cosmic rays up 
to the knee region. There are several ideas to use this background 
for physics relevant for the discussed subject, though corresponding 
activities are not well elaborated. 
Nevertheless the data can be used to improve the results and/or 
to test the interaction models at moderate energies.\\
In order to improve the present situation, the KASCADE experiment 
is being extended (KASCADE-Grande), covering an area of
$600 \times 600\,$m$^2$ by installing the former EAS-TOP
detectors~\cite{grande} at the site of the 
Forschungzentrum Karlsruhe. 
The extended detector installation will 
be able to measure the CR energy spectrum and mass composition 
up to $10^{18}\,$eV in a total measuring time of 3-4 years. 
The data taking will start end of 2002.
The KASCADE-Grande experiment will include the full
information provided by the original KASCADE multi-detector setup 
for each measured single event. The multiparameter set-up will 
enable consistancy studies of the high-energy hadronic interaction 
models. This possibility is of great importance as a step 
towards the development and tests of the high-energy hadronic 
interaction models (see S.~Ostapchenko, these proceedings) used for 
interpreting ultra-high energy cosmic ray observations. \\
The origin of the knee is still a puzzle. But a large progress 
is obvious thanks to larger statistical accuracy of multiparameter 
measurements combined with sophisticated analysis methods. The 
main uncertainty and hence pointing to the key to the solution 
is the missing knowledge of the detailed interaction mechanisms 
for the energy and kinematic ranges under consideration. Efforts in
this direction are a common subject of particle and cosmic ray
physicists~\cite{needs}.
Such efforts will become even more important with CR 
measurements at much higher energies, around the ankle" 
of the CR spectrum and above 
with experiments like the Pierre-Auger-Project, 
HiRes, or EUSO (see H.~Bl\"umer, these proceedings).

\ack
I would like to thank the organizers of the Moscow ECRS for their 
kind invitation and hospitality. I acknowledge fruitful discussions 
and valuable clarifications with my colleagues of KASCADE and  
other experiments.

\end{document}